\documentclass{article}

\PassOptionsToPackage{numbers}{natbib}

    \usepackage[preprint]{neurips_2020}

\usepackage[utf8]{inputenc} 
\usepackage[T1]{fontenc}    
\usepackage{hyperref}       
\usepackage{url}            
\usepackage{booktabs}       
\usepackage{amsfonts}       
\usepackage{nicefrac}       
\usepackage{microtype}      
\usepackage{amsmath}
\usepackage{amssymb}
\usepackage{graphicx}
\usepackage{wrapfig}
\DeclareMathOperator{\atantwo}{atan2}

\title{Low-Bandwidth Communication Emerges Naturally in Multi-Agent Learning Systems}

\author{%
  Niko A.~Grupen\thanks{Correspondence to: \texttt{niko@cs.cornell.edu}}, \, Daniel D. Lee, \, Bart Selman\\
  Department of Computer Science\\
  Cornell University\\
}

\begin{document}

\maketitle

\begin{abstract}
    In this work, we study emergent communication through the lens of cooperative multi-agent behavior in nature. Using insights from animal communication, we propose a spectrum from low-bandwidth (e.g. pheromone trails) to high-bandwidth (e.g. compositional language) communication that is based on the cognitive, perceptual, and behavioral capabilities of social agents. Through a series of experiments with pursuit-evasion games, we identify multi-agent reinforcement learning algorithms as a computational model for the low-bandwidth end of the communication spectrum.
\end{abstract}

\section{Introduction}
Recent work in the multi-agent reinforcement learning (MARL) community has shown that cooperative agents can effectively learn protocols that improve performance on partially-observable tasks \cite{lazaridou2020emergent} and, given additional structural learning biases, exhibit language-like properties (e.g. Zipf's law \cite{chaabouni2019anti} and compositionality \cite{chaabouni2020compositionality, cogswell2019emergence, resnick2019capacity}). Though the study of emergent communication is fundamentally an \textit{ab initio} approach to communication as compared to top-down approaches to language learning \cite{brown2020language, devlin2018bert, vaswani2017attention}, the majority of recent methods target protocols with sophisticated structure and representational capacity, like that of human language \cite{lazaridou2020emergent, lowe2019pitfalls}.

Multi-agent cooperation in nature, however, gives rise to a diverse scope of communication protocols that vary significantly in their structure and the complexity of the information they can convey. In animal communication \cite{bradbury1998principles}, whether intra- or inter-species, a protocol is shaped by the physical capabilities of both the speaker(s) and the listener(s). For example, reef-dwelling fish use a variety of body shakes to communicate \cite{vail2013referential, bshary2006interspecific}, whereas chimps maintain a diverse vocal repertoire \cite{boesch1989hunting}. The diversity of skill found in the animal kingdom rewards a \textit{spectrum of communication} that ranges from low-bandwidth implicit communication (e.g. pheromone trails \cite{beckers1989colony, holldobler1999multimodal}) to rich, high-bandwidth communication (e.g. natural language). If our goal is to endow multi-agent systems with high-bandwidth, language-like communication, it is necessary to first understand the environmental, social, and physical pressures that leads low-bandwidth communication to arise in learning systems.

In this paper, we outline the communication spectrum that exists in nature through a series of examples and identify communication as a system that emerges naturally under optimization pressure. Through experiments in the domain of pursuit-evasion games \cite{isaacs1999differential}, we show that existing MARL algorithms effectively learn low-bandwidth communication strategies.

\section{The communication spectrum in nature}
\begin{figure}[]
    \centering
    \makebox[\textwidth][c]{\includegraphics[width=1.0\textwidth]{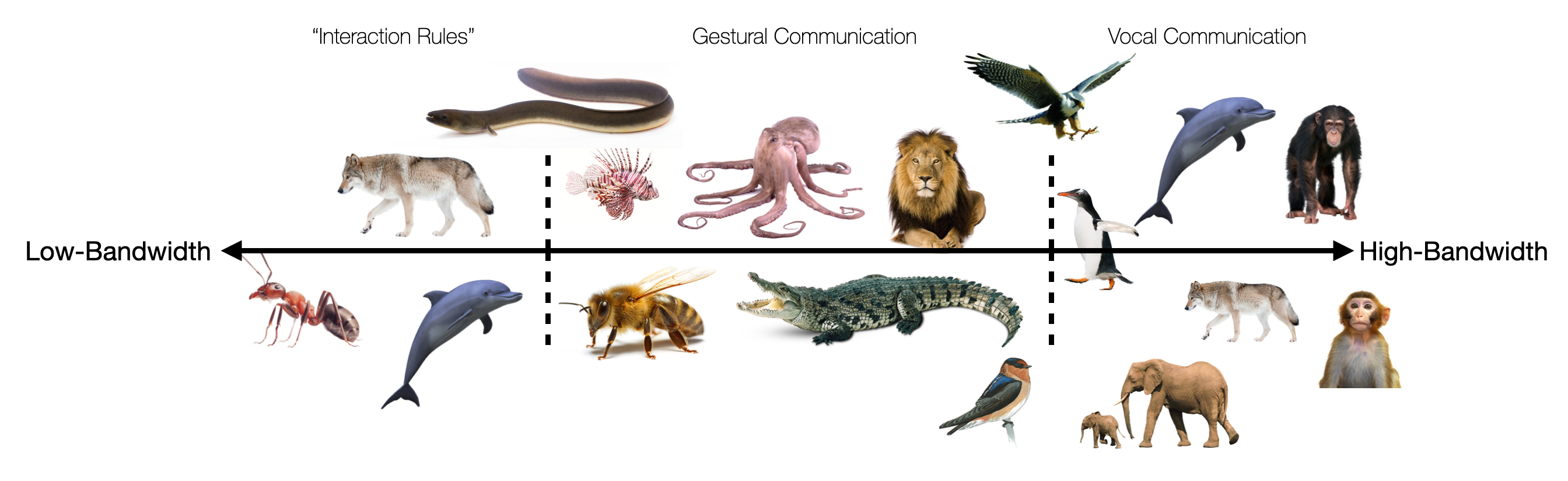}}
    \caption{A snapshot of the communication spectrum. Some animals appear in multiple places in this figure, signifying their use of multiple forms of communication.}
    \label{fig_spectrum}
\end{figure}
Species throughout the animal kingdom leverage communication to achieve efficient social coordination. Here we survey biological examples of communication with the goal of understanding how its emergence can be modeled computationally. Though communication is used in a variety of social contexts, we focus on signals that are produced for the purposes of group foraging and cohesion. A visual depiction of the spectrum is provided in Figure \ref{fig_spectrum}.

\subsection{Communication in the wild}
Fundamentally, communication is an information channel with which animals can coordinate and survive in a partially-observable world. For many species, survival requires finding reliable food sources. Moreover, foraging often involves intra- or inter-species collaboration, leading to ``social predation" \cite{lang2017multidimensional}. Possibly the simplest example of group foraging is that of the Weaver ant, who leaves a trail of pheromones guiding other ants to a stationary food source \cite{beckers1989colony, holldobler1999multimodal}. This is a particularly low-bandwidth medium of communication, as the Weaver ant drops pheromones reflexively, not intentionally. Social animals that rely on capturing mobile food sources---e.g. reef-dwelling fish (grouper \cite{vail2013referential, bshary2006interspecific}, lionfish \cite{lonnstedt2014lionfish}), reptiles (crocodile \cite{dinets2015apparent}), and some mammals (lion \cite{schaller2009serengeti}, chimpanzee \cite{boesch1989hunting})---use gestural, chemical, and postural communication to coordinate group movements during foraging. Species that have evolved to produce sound---such as wolves \cite{peterson2003wolf, herbert2016understanding}, dolphins \cite{quick2012bottlenose}, penguins \cite{choi2017group}, cliff swallows \cite{brown1991food}, and Aplomado falcons \cite{hector1986cooperative}---leverage the high-bandwidth medium that the larynx (or syrinx in avian species) provides by vocalizing the location of food sources. Communication also occurs when coordination involves localizing other pack- or herd-members. In addition to the aforementioned species, whales \cite{mann2000cetacean, whitehead2003sperm}, African elephants \cite{poole1988social}, cape gannets \cite{thiebault2016seabird}, and Rhesus monkeys \cite{mason1962communication} communicate for the purposes of group cohesion.

We find that the sophistication of communication depends heavily on physical capabilities and survival difficulty. Together, these conditions define a spectrum of communicative bandwidth upon which each of these emergent communication protocols falls. While the higher ends of the spectrum begin to resemble language-like communication, the lowest end consists of implicit behavioral information generated through patterns of activity. For example, though wolves and dolphins engage in vocal communication to localize prey, neither species communicates vocally during the foraging act. Instead, they adhere to simple ``interaction rules" in which individual group members adjust their position or orientation based on the positions and orientations of other members of the group \cite{muro2011wolf}. Though this low-bandwidth form of communication differs from explicit symbolic knowledge, it is equally important in understanding the emergence of communication in multi-agent systems.

\subsection{A computational analogue}
Each of the examples in the previous section involves sensorimotor systems that engage in cooperative behavior within their biological constraints. In accordance with signalling theory---which posits that communication is preserved only if all parties benefit from the communicated information \cite{connelly2011signaling}---there appears to be a common mechanism underlying the evolution of communication that is held together by mutual reward (i.e. successful foraging). Computationally, this mechanism has a parallel in reinforcement learning, whose primary objective is to maximize expected reward $J(\phi)$, given by $J(\phi) = \mathop{\mathbb{E}}_{\tau \sim p_\phi(\tau)} [\sum_{t=0}^T \gamma^t r(s_t, a_t)]$ where $s_t$ is an environmental state at time $t$, $a_t$ is an action chosen at time $t$ according to parameters $\phi$, and $p_\phi(\tau)$ is a probability distribution over a trajectory $\tau = \{s_0, a_0, ..., s_T, a_T \}$, $r$ is a reward function indicating the strength or weakness of selecting action $a_t$ in state $s_t$, and $\gamma \in [0,1]$ is a discount factor. The connection between reward maximization and action-space communication has been identified in prior work \cite{mordatch2017emergence, baker2019emergent}. In the next section, we explore this connection further and use a variant of this learning paradigm to investigate whether low-bandwidth ``interaction rules" emerge naturally amongst artificial agents.

\section{Learning low-bandwidth communication}
\label{sec_learning}
We identify inferred behavioral communication as the first step towards a foundational account of emergent communication. We hypothesize that learning agents will naturally develop ``interaction rules" and, in turn, outperform methods that do not leverage low-bandwidth communication. To test this hypothesis, we define a set of experiments in the domain of pursuit-evasion games \cite{isaacs1999differential}. In this section, we describe our experimental domain, approach to multi-agent learning, and results against non-communicative baselines. Please see the Appendix for additional details.

\subsection{Experimental setup}
\label{sec_learning_exp}
We consider a pursuit-evasion game in $\mathbb{R}^2$ between $N$ predators $\textbf{P} = \{P_1, ..., P_N \}$ and a single prey $E$. Each agent $i \in \{\textbf{P}, E\}$ is defined by a state $q_i = [x_i \, y_i \, \theta_i]^T$, representing its position and heading at time $t$. Movement of each agent is described as $\dot{q_i} = [\dot{x}_i \, \dot{y}_i \, \dot{\theta_i}]^T =  [\Vec{v}_i \cos(\theta_i)\ \ \Vec{v}_i\sin(\theta_i)\ \ \tan^{-1}(\Vec{v}_i)]^T$, where $\Vec{v}_i \in \{\Vec{v}_{P_1}, ..., \Vec{v}_{P_N}, \Vec{v}_E\}$ is the agent's velocity. The goal of $\textbf{P}$ is to capture $E$ as quickly as possible, where capture is defined as a collision between predator and prey. The game is terminated when the prey is caught (predator victory) or the maximum number of time-steps is reached (prey victory). 

To simulate our experiments, we use a modified version of the pursuit-evasion environment introduced by \citet{lowe2017multi}. First, we project the planar environment onto a torus. In unbounded planar pursuit-evasion, the prey has a significant advantage in the $\lvert \Vec{v}_P \rvert < \lvert \Vec{v}_E\rvert$ case, as it can outrun the predators in any direction. Toroidal pursuit-evasion forces interaction between the agents, as the prey cannot permanently escape. Next, we remove all of the constraints on agent motion---enabling instantaneous change of velocities---and remove any obstacles from the environment. These adjustments increase the difficulty of the task, as predators cannot rely on changing the direction of the prey to slow it down or pinning it against an obstacle. In general, the game as we have defined it is easily solved when $\lvert \Vec{v}_P \rvert > \lvert \Vec{v}_E\rvert$. The predators can pursue the prey greedily in a straight-line chase. When $\lvert \Vec{v}_P \rvert \leq \lvert \Vec{v}_E\rvert$, however, we can define a prey strategy of near impossible difficulty to an uncoordinated group of predators.

\subsection{Training details}
\begin{figure}[]
    \centering
    \makebox[\textwidth][c]{\includegraphics[width=1.0\textwidth]{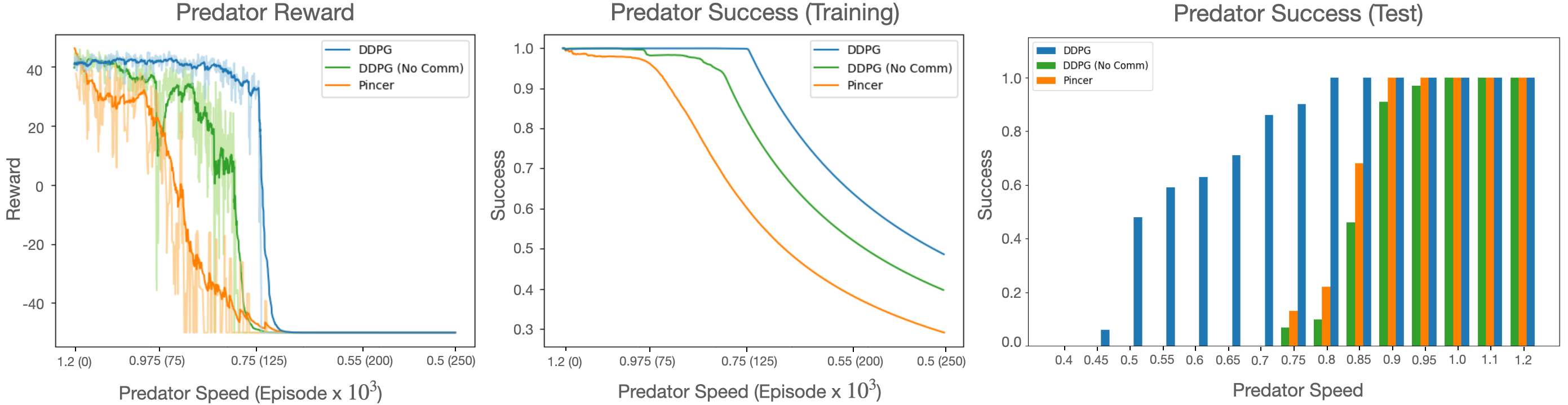}}
    \caption{(a) Predator reward during training (see Appendix \ref{sec_apdx_reward} for definition of reward). (b) Capture success of the predators during training. Note that $\lvert \Vec{v}_P \rvert$ decays throughout training, so maintaining reward/capture success is indicative of improved predator coordination. Less sophisticated methods (i.e. DDPG (No Comm) and Pincer) falter as $\lvert \Vec{v}_P \rvert$ decreases. (c) Capture success at test-time for both learned and potential field predator policies. The difference between (b) and (c) is that during training the network tunes its parameters consistently as $\lvert \Vec{v}_P \rvert$ decreases---which is why performance rebounds and stays near perfect---whereas during test we decay $\lvert \Vec{v}_P \rvert$ for a fixed policy that was previously trained until $\lvert \Vec{v}_P \rvert = 0.6$. Test performance therefore drops for lower velocities.}
    \label{fig_results}
\end{figure}
Each predator $P_i$ is initialized with a deterministic action policy $\mu_{\phi_{P_i}}(s_i)$ that is parameterized by a neural network. Each agent receives a complete observation of the environment state $s_t = \{q_{P_1}, ..., q_{P_N}, q_E \}$ and outputs velocity commands as an action. Action policies $\{\mu_{\phi_{P_1}}, ... ,\mu_{\phi_{P_N}}\}$ are trained in a decentralized manner, following the Deep Deterministic Policy Gradients (DDPG) algorithm \cite{lillicrap2015continuous}. To aid the predators during training, we introduce curriculum learning using velocity bounds. Specifically, we start training with $\lvert \Vec{v}_P \rvert > \lvert \Vec{v}_E\rvert$ and anneal it slowly over time until $\lvert \Vec{v}_P \rvert < \lvert \Vec{v}_E\rvert$ by some threshold ($\lvert \Vec{v}_P \rvert = 0.6 \cdot \lvert \Vec{v}_E\rvert$ in our experiments). This curriculum effects both the reward and capture success during training, as shown in Figure \ref{fig_results}(a) and (b). As $\lvert \Vec{v}_P \rvert$ decays, the predators must learn a more sophisticated cooperative pursuit strategy.

\paragraph{Prey escape strategy}
We define a potential field control policy for the prey, which minimizes the following cost function:
\begin{equation}
    U(\theta_E) = \sum_i \bigg(\frac{1}{r_i}\bigg) \cos(\theta_E - \theta_i)
\end{equation}
where $r_i = d(q_E, q_{P_i})$ is the distance between the prey's location $q_E$ and the location of the $i$-th predator $q_{P_i}$ and $\theta_i = \atantwo(y_{P_i}, x_{P_i})$ is the relative angle. Intuitively, this objective function incentivizes the prey to avoid capture, encouraging it to move towards the bisector of two predators, while repelling it from any one predator.

\paragraph{Predator baselines}
\label{sec_training_baseline}
At each time-step, the prey will choose the heading $\theta_E^* = \min_{\theta_E}[U(\theta_E)]$ that minimizes its cost function. As a baseline for the predators, we define a potential field function that exploits knowledge of the prey's objective:
\begin{equation}
    \label{eq_maxmin}
    F(\boldsymbol{\theta_i, r_i}) = \underset{\boldsymbol{\theta_i, r_i}}{\max} \big[ \underset{\theta_E}{\min} \big[ U(\theta_E)\big] \big] = \underset{\boldsymbol{\theta_i, r_i}}{\max} \bigg[ \underset{\theta_E}{\min} \bigg[ \sum_i \bigg(\frac{1}{r_i}\bigg) \cos(\theta_E - \theta_i) \bigg] \bigg]
\end{equation}
\noindent where $\boldsymbol{\theta_i}$ and $\boldsymbol{r_i}$ are the sets of distances and headings, respectively, for each predator location relative the prey. By maximizing the prey's objective, the predators are incentivized to surround the prey equidistantly and prevent it from escaping along a bisector. This produces an encircling behavior similar to the predation strategies found in wolf and dolphin groups (see top rows in Figure \ref{fig_qual_results}(b) in Appendix \ref{sec_apdx_qual}). For this reason, we refer to the potential-field strategy as the ``pincer". Crucially, this hand-crafted system does not support communication in the form of ``interaction rules" as the predators close in on the prey. We also compare to a non-communicative variant of DDPG. In particular, we prevent each predator from observing its fellow teammates, thereby removing their ability to coordinate.

\subsection{Results}
\label{sec_results}
We evaluate performance of both the learned and potential field predators as a function of the velocity advantage of the prey. Results are provided in Figure \ref{fig_results}(c) for a variety of  $\lvert \Vec{v}_P \rvert$ values. Both strategies perform well in the $\lvert \Vec{v}_P \rvert > \lvert \Vec{v}_E\rvert$ cases, as expected, but the DDPG predators significantly outperform the baseline predators as $\lvert \Vec{v}_P \rvert$ decays. This verifies that simple coordination (e.g. encircling) is not enough to capture a sophisticated prey---an additional information exchange is required. We posit that the ability of the predators to communicate implicitly through physical ``interaction rules" is key to their success at lower velocities. Through additional qualitative analysis (Appendix \ref{sec_apdx_qual}), we show that the DDPG predators may indeed utilize a low-bandwidth form of communication in which they adaptively modify their position based on the movements of other predators. This behavior is similar to the ``interaction rules" displayed by dolphins and wolves during foraging.

\section{Conclusion}
We explored the spectrum of communication that exists in nature and introduced low-bandwidth communication as a foundation for robust emergent communication. Experimentally, we showed that low-bandwidth ``interaction rules" emerge naturally from MARL systems, resulting in increased capture success in a pursuit-evasion game. In future work, we will continue to study how common principles can contribute to integrated, communicative behavior. We will also examine how low-bandwidth communication evolves if the agents are exposed to imperfect state information.

\begin{ack}
    We thank the reviewers for their valuable feedback. This research was supported by NSF awards CCF-1522054 (Expeditions in computing), AFOSR Multidisciplinary University Research Initiatives (MURI) Program FA9550-18-1-0136, AFOSR FA9550-17-1-0292, AFOSR 87727, ARO award W911NF-17-1-0187 for our compute cluster, and an Open Philanthropy award to the Center for Human-Compatible AI.
\end{ack}

\bibliography{bib}
\bibliographystyle{unsrtnat}

\appendix
\section{Background}
\label{sec_apdx_background}
Here we provide brief descriptions of concepts that are useful for understanding of experimental setup.

\subsection{Partially-observable Markov games}
In addition to the environment dynamics outlined in Sec. \ref{sec_learning_exp}, our game is defined by action spaces $\boldsymbol{\mathcal{A}} = \{\mathcal{A}_{P_1}, ... , \mathcal{A}_{P_N}, \mathcal{A}_E\}$ and observation spaces $\boldsymbol{\mathcal{O}} = \{\mathcal{O}_{P_1}, ... , \mathcal{O}_{P_N}, \mathcal{O}_E\}$ for each of the agents. Each agent $i$ is initialized with a deterministic action policy $\mu_{\phi_i}(s_i)$. Upon selecting a set of actions $\{a_{P_1}, ..., a_{P_N}, a_E \}$, the environment responds by transitioning from its current state $s_t \sim \mathcal{S}$ to a new state $s_{t+1} \sim \mathcal{S}$, as governed by the transition function $T:\mathcal{S} \times \mathcal{A}_{P_1} \times ... \times \mathcal{A}_{P_N} \times \mathcal{A}_E \rightarrow \mathcal{S}$, where $\mathcal{S}$ is a state space representing all possible configurations of our $N+1$ agents. The environment also produces a reward $r: \mathcal{S} \times \mathcal{A}_{P_1} \times ... \times \mathcal{A}_{P_N} \times \mathcal{A}_E \rightarrow \mathbb{R}$ indicating the strength or weakness of each agent’s decision-making. The goal of each agent $i$ is to maximize its expected return $R_i = \sum_{t=0}^T \gamma^t r_i^t$ over some time horizon $T$. This formulation is consistent with the partially-observable Markov game framework \cite{littman1994markov}, which itself is a variant of the classical Markov decision process (MDP).

\section{Additional experimental details}
\label{sec_apdx_details}
In this section, we present additional details for our pursuit-evasion experiments.

\begin{figure}[b!]
    \centering
    \makebox[\textwidth][c]{\includegraphics[width=1.0\textwidth]{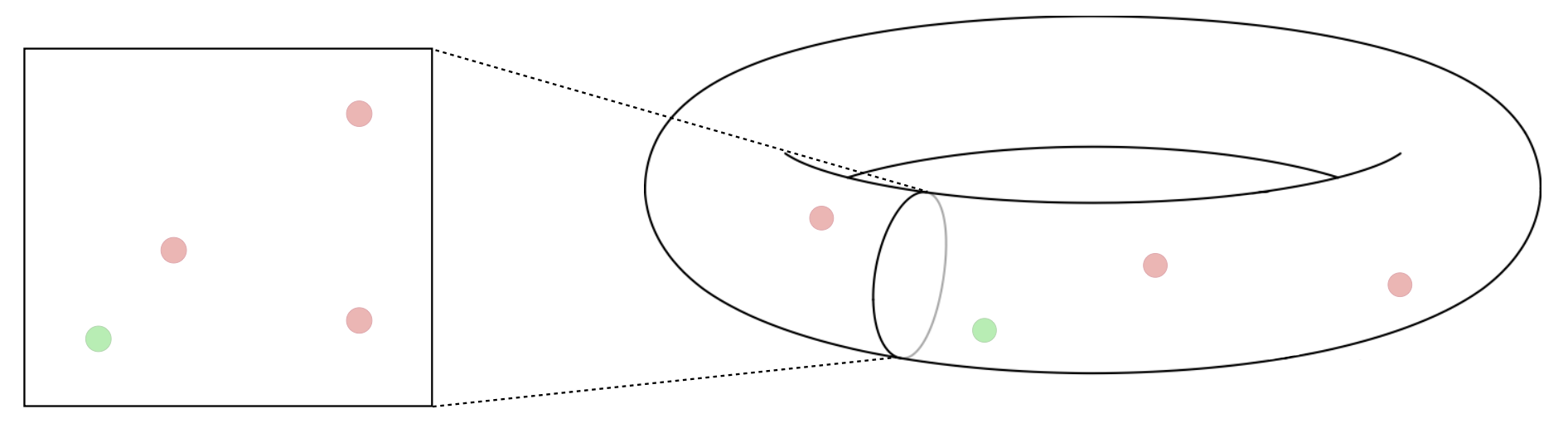}}
    \caption{A projection of the planar pursuit-evasion game onto a torus.}
    \label{fig_torus}
\end{figure}

\subsection{Pursuit-evasion on a torus}
As discussed in Sec. \ref{sec_learning_exp}, we modify the environment by placing the planar pursuit-evasion environment on a torus. This amounts to connecting each horizontal and vertical edge with its opposite counterpart. Visually, this means that each agent, upon moving across the boundary of the visible plane, will reappear on the other side of the plane. A visualization of this projection is shown in Figure \ref{fig_torus}.

\subsection{Game initialization and reward structure}
\label{sec_apdx_reward}
\begin{figure}[]
    \centering
    \makebox[\textwidth][c]{\includegraphics[width=1.0\textwidth]{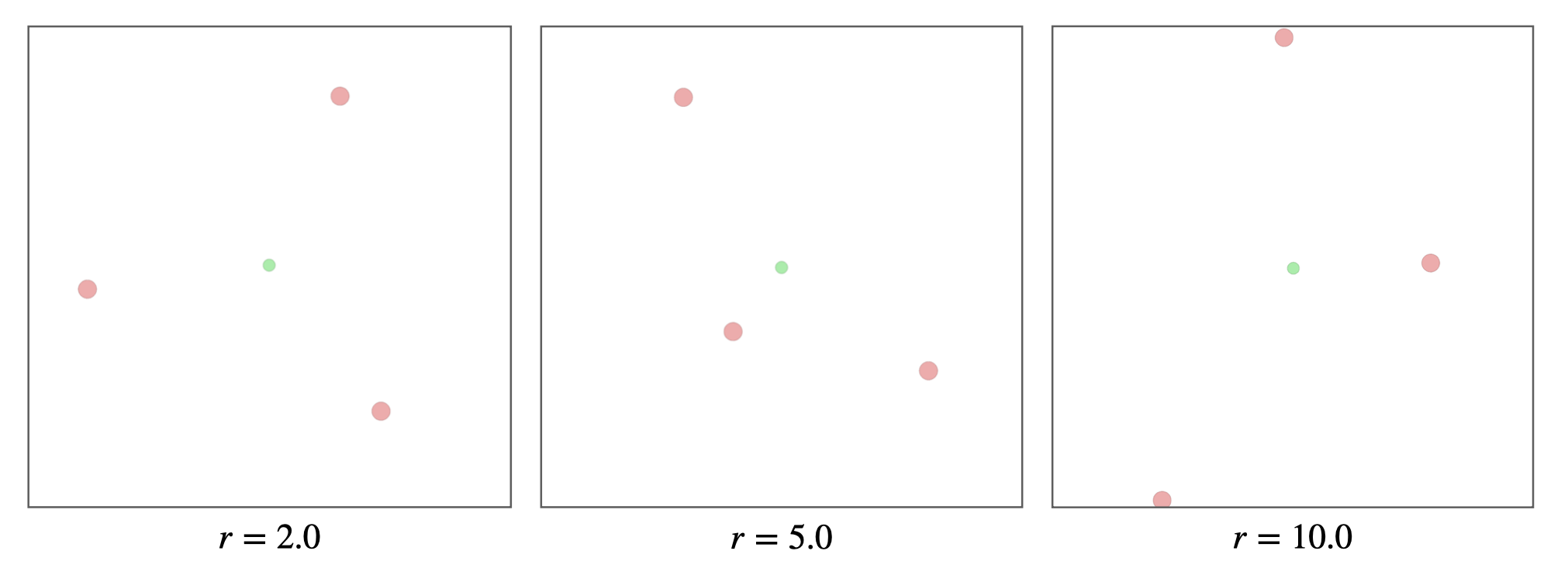}}
    \caption{A snapshot of the pursuit-evasion environment with $N = 3$ predators.}
    \label{fig_env}
\end{figure}

The predators are initialized in a circular formation of radius $r$ around the prey, as shown in Figure \ref{fig_env}. The prey is initially centered at the origin, encircled (in toroidal coordinates) by the initial predator formation.

The predators' reward function is structured as follows:
\begin{equation*}
    r =
    \begin{cases}
    50.0 & \textrm{if prey captured}
    \\
    -0.1, & \textrm{otherwise}
    \end{cases}
\end{equation*}
\noindent where capture is defined as a collision between predator and prey. The small negative penalty incentives the predators to catch the prey quickly. The simulation runs for a maximum of $500$ time-steps, yielding a minimum total reward of $-50.0$ per episode.

\subsection{Curriculum learning}
\label{sec_apdx_curriculum}
Curriculum learning \cite{bengio2009curriculum} is a useful technique for speeding up the rate at which RL systems learn, especially when rewards are sparse. In our experiments, the predators do not receive a positive reward signal unless the prey is caught. Due to the sophistication of the prey policy, the likelihood of randomly initialized action policies capturing the prey is extremely low when $\lvert \Vec{v}_P \rvert \leq \lvert \Vec{v}_E\rvert$. To help the predators experience reward signal early in the training process, we initialize $\lvert \Vec{v}_P \rvert > \lvert \Vec{v}_E\rvert$ and decay predator velocity linearly over time.

\subsection{Prey escape strategy (cont.)}
The goal of the prey strategy is to define a potential field in $\theta$-space such that the prey naturally moves towards the maximum bisector between two predators. Given predator positions $\{q_{P_1}, ..., q_{P_N}\}$ in prey-centric coordinates, we compute polar coordinates:
\begin{align*}
    r_i &= d(q_E, q_{P_i})
    \\
    \theta_{P_i} &= \textrm{atan2}(y_{P_i}, x_{P_i})
\end{align*}

\noindent for each predator $P_i$ relative the prey. Next, we use the relative angles of the predators to define a potential field that will push the prey towards a bisector:
\begin{equation*}
    U(\theta_E) = \sum_i \cos(\theta_E - \theta_{P_i})
\end{equation*}

\noindent Using Ptolemy's difference formula, we can expand the potential field as:
\begin{align*}
    U(\theta_E) &= \sum_i \cos(\theta_E - \theta_{P_i})
    \\
    &= \sum_i \cos(\theta_E)\cos(\theta_{P_i}) + \sin(\theta_E)\sin(\theta_{P_i})
    \\
    &= A \cos(\theta_E) + B \sin(\theta_E)
\end{align*}
\noindent when we plug-in the known $\theta_{P_i}$ values. The function $U(\theta_E)$ is maximized/minimized for values of $A$ and $B$ such that:
\begin{equation*}
    \nabla U(\theta_E) = -A\sin(\theta_E) + B\cos(\theta_E) = 0
\end{equation*}
\noindent which results in:
\begin{align*}
     B\cos(\theta_E) &= A\sin(\theta_E)
     \\[6pt]
     \tan(\theta_E) &= \frac{B}{A}
\end{align*}
\noindent We select the prey's next heading by following the direction of the negative gradient ($-\nabla U(\theta_E)$) and pursue it at maximum speed. Further, modulating the cost function by $r_i$:
\begin{equation*}
    U(\theta_E) = \sum_i \bigg(\frac{1}{r_i}\bigg) \cos(\theta_E - \theta_{P_i})
\end{equation*}
\noindent allows the prey to modify its bisector based on the distance to each predator. This helps significantly when the prey is stuck in symmetric formations.

\subsection{Baseline predator strategy (cont.)}

\begin{wrapfigure}{r}{0.45\textwidth}
    \begin{center}
    \includegraphics[width=0.42\textwidth]{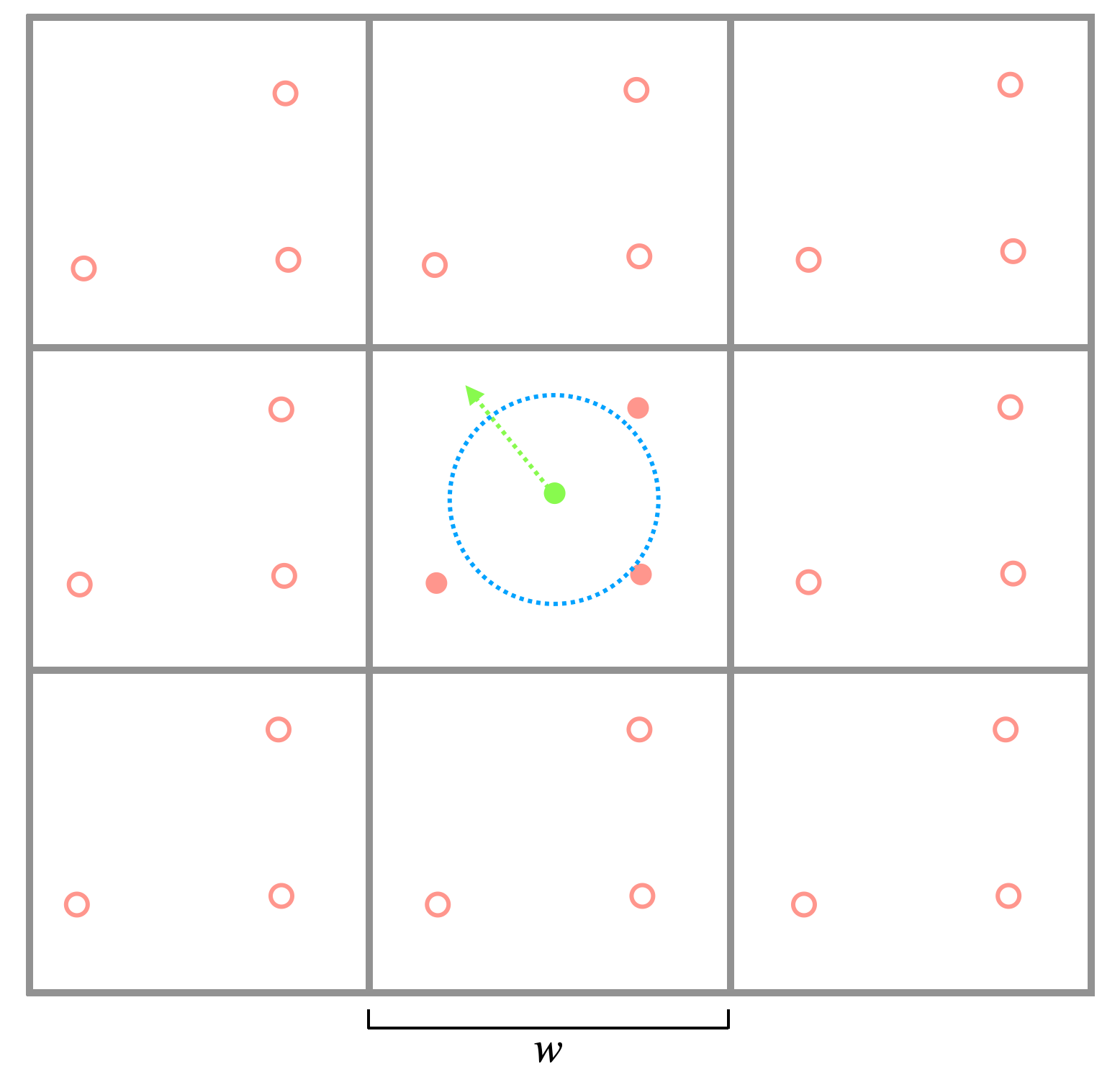}
    \caption{The torus environment unrolled $k=1$ times in each direction. The filled in red circles denote the ``active" predators that are pursuing the prey at the current time-step, while the empty circles. We consider only a single prey, located in the center tile.}
    \label{fig_replica}
    \end{center}
\end{wrapfigure}
The potential field approach described in Section \ref{sec_training_baseline} requires optimizing over both $\boldsymbol{\theta_i}$ and $\boldsymbol{r_i}$. Fortunately, we can exploit the toroidal structure of the environment to construct an optimization routine that solves for $\boldsymbol{\theta_i}$ and $\boldsymbol{r_i}$ discretely. Starting from the planar projection in Figure \ref{fig_replica}, unrolling the torus $k$ steps in each direction generates $(2k + 1)^2$ replications of the current environment state. Rather than solving for optimal $\boldsymbol{\theta_i}$ and $\boldsymbol{r_i}$ values directly, we find the set $\boldsymbol{P}$ of predators that maximize Eqn.\eqref{eq_maxmin} across all replications of the environment. We constrain the problem by limiting selections of each predator $P_i$ to replications of \textit{itself only}. This dramatically cuts down the number of possible sets $\boldsymbol{P}$ from $\binom{(2k+1)^2N}{N}$ to $\binom{(2k+1)^2}{1} \cdot \binom{(2k+1)^2}{1} \cdot \binom{(2k+1)^2}{1}$, where $N$ is the number of predators in the environment. Thus, we solve Eqn.\eqref{eq_maxmin} via a discrete optimization over each of the $((2k + 1)^2)^3$ possible predator selections.

The resulting set $\boldsymbol{P}$ defines the set of ``active" predators that will pursue the prey directly at the next time-step. Due to the nature of the prey's objective function---it is attracted to bisectors and repulsed from predators---the maximum $\boldsymbol{P}$ tends to favor symmetric triangular formations. Though this method obviously does not scale well with $N$ and $k$, we found that we are able to find a sufficient maximizer with low values of $k$ (i.e. $k=1$ in our experiments). The replication process is shown for the $k=1$ case in Figure \ref{fig_replica}. Note that we discriminate between ``active" predators---i.e. those $P_i \in P$ pursuing the prey at the current time-step---from ``inactive" predators.

\section{Qualitative results}
\label{sec_apdx_qual}
\begin{figure}[]
    \centering
    \makebox[\textwidth][c]{\includegraphics[width=1.0\textwidth]{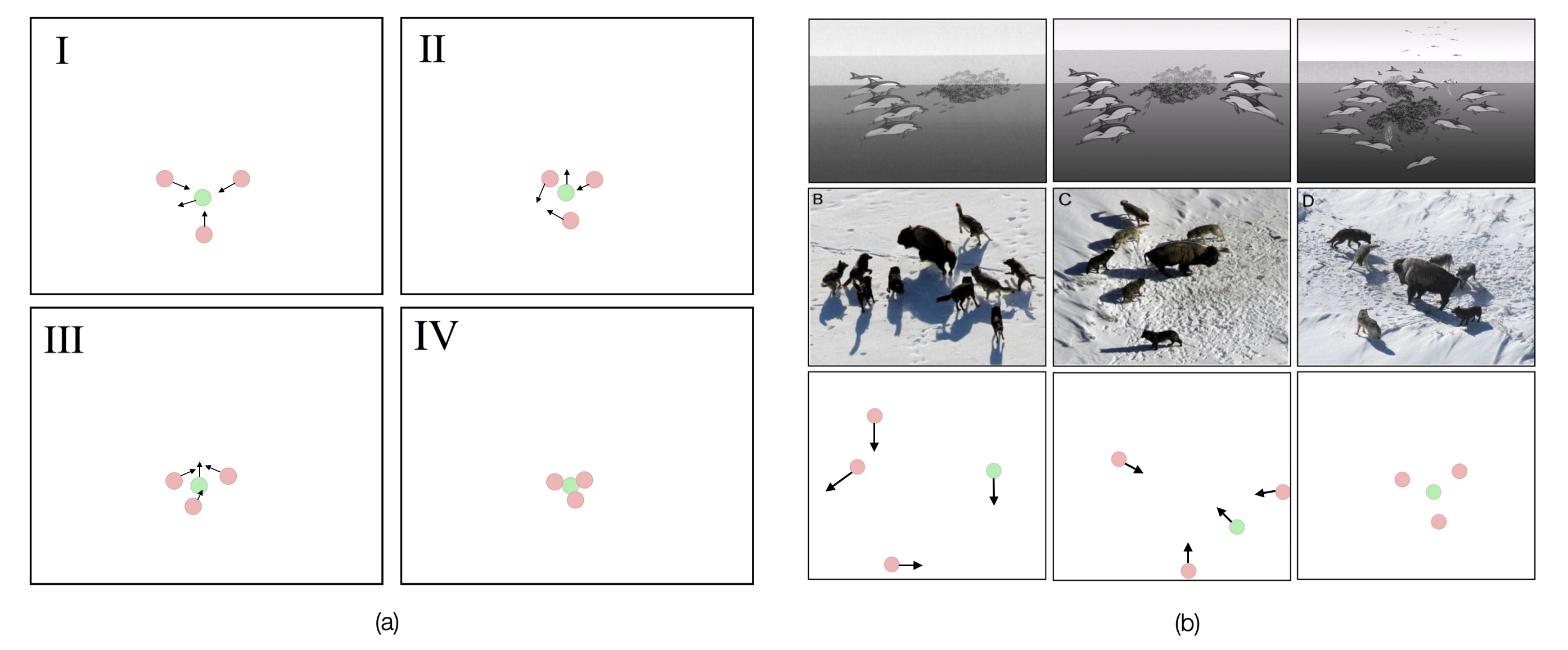}}
    \caption{Qualitative results from the pursuit-evasion experiment. (a) The predators coordinate to capture the prey, displaying positional shifts similar to low-bandwidth ``interaction rules". (b) \textit{Top:} A diagram of dolphin foraging strategies documented in \cite{neumann2003feeding}. \textit{Middle:} Photos of wolves coordinating while hunting, as shown in \cite{muro2011wolf}. \textit{Bottom:} The learned behavior of our multi-agent system.}
    \label{fig_qual_results}
\end{figure}
In addition to the results presented in Section \ref{sec_results}, we perform post-hoc analysis of predator trajectories as they pursue and encircle the prey. Example trajectories are shown in Figure \ref{fig_qual_results}. By analyzing predator trajectories during pursuit, we find evidence that low-bandwidth communication emerges naturally from MARL algorithms. Not only does the pursuit strategy learned by the agents mimic the foraging behaviors of the animals we have studied thus far, but it also displays low-bandwidth communication (e.g. ``interaction rules"). In particular, the predators appear to adjust their position slightly in response to the movements of fellow predators as they close in on the prey (see Figure \ref{fig_qual_results}(a)). Moreover, the predators occasionally move away from the prey---something a less coordinated strategy would not do---to maintain the integrity of the group formation. This could partially explain the performance differential between the DDPG predators and the potential field predators, as the potential field predators have no basis for making small-scale adaptive movements. Though these results are only qualitative to this point, they are encouraging examples of emergent low-bandwidth communication.

\end{document}